\documentclass[manuscript]{aastex}

\shorttitle{Further constraints on the optical transmission spectrum of HAT-P-1b}
\shortauthors{Montalto et al.}

\begin{document}

\title{Further constraints on the optical transmission spectrum of HAT-P-1b}

\author{M. Montalto}
\affil{Instituto de Astrof\'isica e Ci\^encias do Espa\c{c}o, Universidade do Porto, CAUP, Rua das Estrelas, PT4150-762 Porto, Portugal}
\email{Marco.Montalto@astro.up.pt}

\author{N. Iro}
\affil{Theoretical Meteorology group Klimacampus, University of Hamburg Grindelberg 5, 20144 Hamburg Germany}

\author{N. C. Santos\altaffilmark{1}}
\affil{Instituto de Astrof\'isica e Ci\^encias do Espa\c{c}o, Universidade do Porto, CAUP, Rua das Estrelas, 4150-762 Porto, Portugal}

\author{S. Desidera}
\affil{INAF - Osservatorio Astronomico di Padova, Vicolo dell’Osservatorio 5, Padova, IT-35122}

\author{J. H. C. Martins\altaffilmark{1},\altaffilmark{2}}
\affil{Instituto de Astrof\'isica e Ci\^encias do Espa\c{c}o, Universidade do Porto, CAUP, Rua das Estrelas, 4150-762 Porto, Portugal}

\author{P. Figueira\altaffilmark{1}}
\affil{Instituto de Astrof\'isica e Ci\^encias do Espa\c{c}o, Universidade do Porto, CAUP, Rua das Estrelas, 4150-762 Porto, Portugal}

\author{R. Alonso\altaffilmark{3}}
\affil{Instituto de Astrof\'\i sica de Canarias, E-38205 La Laguna, Tenerife, Spain}

\altaffiltext{1}{Departamento de F\'isica e Astronomia, Faculdade de Ci\^encias, Universidade do Porto, Rua do Campo Alegre 687, PT4169-007 Porto, Portugal}
\altaffiltext{2}{European Southern Observatory, Alonso de Cordova 3107, Vitacura Casilla 19001, Santiago 19, Chile}
\altaffiltext{3}{Dpto. de Astrof\'isica, Universidad de La Laguna, 38206 La Laguna, Tenerife, Spain\label{La Laguna}}

\begin{abstract}
We report on  novel observations of HAT-P-1 aimed  at constraining the
optical  transmission spectrum  of  the atmosphere  of its  transiting
Hot-Jupiter  exoplanet.   Ground-based differential  spectrophotometry
was performed over two  transit windows using the DOLORES spectrograph
at the  Telescopio Nazionale Galileo (TNG). Our  measurements imply an
average     planet     to    star     radius     ratio    equal     to
R$\rm_p$/R$_*$=(0.1159$\pm$0.0005). This result is consistent with the
value obtained  from recent near infrared measurements  of this object
but differs from previously  reported optical measurements being lower
by  around 4.4  exoplanet scale  heights.  Analyzing  the data  over 5
different  spectral bins  $\sim$600\AA$\,$ wide  we observed  a single
peaked spectrum (3.7 $\sigma$ level) with a blue cut-off corresponding
to  the blue  edge  of the  broad  absorption wing  of  sodium and  an
increased absorption  in the region in  between 6180-7400$\rm\AA$.  We
also infer that the width of  the broad absorption wings due to alkali
metals  is likely  narrower than  the one  implied by  solar abundance
clear atmospheric  models.  We interpret  the result as  evidence that
HAT-P-1b has a partially  clear atmosphere at optical wavelengths with
a more  modest contribution from  an optical absorber  than previously
reported.
\end{abstract}

\keywords{Techniques: spectroscopic --- Planets and satellites: atmospheres --- Planets and satellites: individual: HAT-P-1b}

\section{Introduction}

\noindent
The  possibility to characterize  transiting exoplanet  atmospheres by
means  of  high  precision  spectrophotometric  measurements  acquired
during  transits was  theorized  in  a few  works  early this  century
(Seager \& Sasselov 2000; Brown 2001).  The light of the hosting star,
filtered  by  the planeary  atmosphere  in  the  terminator region  is
absorbed and scattered by chemical compounds so that, once observed at
different wavelenghts,  the atmosphere may appear more  or less opaque
to  the observer.   Hence,  transit depth  determinations obtained  in
different spectral domains can be used to reconstruct a low resolution
spectrum of the planetary atmosphere.

\noindent
Models  predict that  atmospheres  of cloud  free hot-jupiter  planets
should be dominated in the optical by broad absorption features of the
alkali  metals Na  and  K.  Studies  so  far have  revealed a  certain
diversity  of  atmospheric features.   Sodium  was  first detected  by
Charbonneau  et al.~(2002) in  the atmosphere  of HD209458b  using the
STIS  spectrograph on  board of  HST and  later on  confirmed  also by
ground based  observations (Snellen et  al.~2008).  Subsequent studies
have revealed the presence of alkali metals also in the atmospheres of
other  planets, for  example sodium  was detected  in WASP-17b  (e. g.
Wood et al.  2011; Zhou \&  Bayliss 2012) and HAT-P-1b (Nikolov et al.
2014) while potassium has been  detected in XO-2b (Sing et al.  2011),
WASP-31b (Sing et  al.  2015) and HAT-P-1b (Wilson  et al.  2015). The
atmosphere of other planets seem to be obscured by clouds or hazes, as
in  the case  of WASP-12b  (Sing  et al  2013) and  HD189733 (Pont  et
al. 2013),  albeit for  this planet narrow  line cores of  sodium were
also  found (Redfield  et al.  2008; Huitson  et al.  2012;  Jensen et
al. 2012).

\noindent
Near infrared studies have soon complemented analysis conducted in the
optical  domain, particularly exploiting  the Spitzer  Space Telescope
and lead to  the conclusion that close-in extrasolar  planets could be
broadly subdivided in two big categories according to the structure of
their temperature-pressure  profile (Hubeny  et al.  2003;  Fortney et
al.  2006, 2008; Burrows et al. 2007, 2008). The pM class planets show
a high altitude temperature inversion in their atmospheres whereas the
pL  class planets  do not,  resulting in  different  emergent spectral
energy  distributions  in  the  near  infrared  domain  and  different
spectral signatures (like the  flipping of water bands from absorption
to emission  passing from non inverted to  inverted atmospheres).  The
origin of such  inversion is identified with the  absorption of strong
stellar irradiation  in the  optical domain by  a variety  of possible
absorbers  (Fortney  et al.   2010,  Burrows  et  al 2007,  Zahnle  et
al. 2009).

\noindent
The present study  is focussed on HAT-P-1b, one  interesting member of
the family  of hot-jupiter  planets.  HAT-P-1b is  a low  mean density
($\rho\rm_p\,\sim\,$0.35     g      cm$^{-3}$)     giant     exoplanet
(R$\rm_p\,\sim\,$1.2  R$_J$) discovered by  the HATNet  transit survey
(B\'akos et al.  2007).  The host star ($V=$10.4)
is  a  member  of a  visual  binary  system  with the  two  components
separated by around 11$^{\arcsec}$ on the sky ($\sim\,1500$ AU).  Both
are  G0V stars  and  the companion  star  is around  half a  magnitude
brighter than  the target in the  $V$ filter.  The low  density of the
exoplanet,  the brightness of  the host  star, and  the presence  of a
close-by stellar companion make HAT-P-1b an ideal target for follow-up
studies  aiming at  characterizing its  atmosphere  using transmission
spectroscopy.  In particular, the  close stellar companion can be used
to correct for systematic effects performing simultaneous differential
spectrophotometric measurements of the two stars.

\noindent
HAT-P-1b  is an  interesting object  also in  the context  of inverted
atmosphere theory.   The pM/pL transition is set  in correspondence of
the TiO condensation  limit (Fortney et al. 2008),  and HAT-P-1b would
sit  just at  the  lower edge  of  this limit  providing an  important
benchmark   test   for  theory.    Todorov   et  al.~(2010)   obtained
Spitzer/IRAC  photometry  for  this  object  and  concluded  that  the
observations  were  best fitted  using  an  atmosphere  with a  modest
temperature  inversion, and  that the  planet has  a  strong day/night
contrast ratio implying little energy redistribution.

\noindent
Three studies have been presented  so far discussing the atmosphere of
HAT-P-1b   by  means  of   transmission  spectroscopy.    Wakeford  et
al.~(2013)  presented HST  Wide  Field Camera  3 (WFC3)  near-infrared
transmission spectroscopy detecting a significant absorption above the
5-$\sigma$ level  matching the 1.4  $\mu$m water absorption  band, and
concluding  that a  1000 K  isothermal model  or  a planetary-averaged
temperature   model  best  matched   the  observations.    Nikolov  et
al.~(2014,  hereafter  N14)  presented HST/STIS  optical  transmission
spectroscopy  implying a  strong absorption  shortward of  5500\AA and
possibly  redward  of 8500\AA,  as  well  as  sodium absorption  at  a
3.3$\sigma$   level,  but  no   evidence  for   potassium  absorption.
Interestingly, the  same authors observed  that STIS and  WFC3 spectra
differ  significantly in absolute  radius level  (4.3$\pm$1.6 pressure
scale  heights).  On  the basis  of their  analysis N14  suggested the
presence  of a strong  $extra$-optical absorber  and of  a super-solar
(factor  10$^3$)  sodium to  water  abundance  ratio  to explain  this
result.   This absorber  would partially  block the  sodium  wings and
completely  mask  the  potassium  feature,  yet  leaving  immuted  the
1.4$\mu$m  absorption  band.   Later  Wilson  et  al.~(2015)  detected
potassium in  the atmosphere of this  planet (albeit at  a much higher
resolution than N14) using the GTC telescope in narrowband imaging and
employing the flexible red  tunable filters technique.  To explain the
amplitude  of  the  detected  feature the  authors  hypothesized  that
high-temperatures at the base of the upper atmosphere may result in an
increased scale  height, and/or suggested that a  lower mean molecular
weight  could  be present  caused  by  the  dissociacion of  molecular
hydrogen into atomic hydrogen by EUV flux from the host star.

\noindent
In this  work we present novel  ground-based transmission spectroscopy
measurements for two full transit events of HAT-P-1b obtained with the
Telescopio Nazionale Galileo (TNG) and the low resolution spectrograph
DOLORES (Device Optimized for the  LOw RESolution). Our data cover the
spectral range  between $\sim$4900\AA$\,$ and  $\sim$8000\AA, and were
obtained between August and October 2012.

\noindent
Our  observations  are  presented in  Sect.\ref{sec:observations},  in
Sect.\ref{sec:data_analysis} we review our data analysis technique, in
Sect.\ref{sec:lightcurve_analysis} we present the study of the transit
lightcurves, in  Sect.\ref{sec:results} we discuss the  results and in
Sect.\ref{sec:conclusions} we summarize and conclude.

\section{Observations}
\label{sec:observations}

We observed  two full transits of  HAT-P-1b in August 14,  2012 and in
October  20, 2012.   The data  were  collected with  DOLORES, the  low
resolution spectrograph and camera installed at the Nasmyth B focus of
the TNG.  The  camera is equipped with a 2100 x  2100 E2V 4240 Thinned
back-illuminated, deep-depleted, Astro-BB coated CCD with a pixel size
of 13.5  $\mu$m corresponding  to a pixel  scale of  0.252 arcsec/pix.
The CCD  gain factor  is 0.97 e$^{-}$/ADU  while the readout  noise is
around 9 e$^{-}$  RMS.  We used the LR-B  grism providing a dispersion
of 2.52 \AA/pix and adopted a slit width of 10 arcsec.  The resolution
was therefore  seeing limited.   An average seeing  of 0.8  arcsec was
measured at the center of our spectral window implying a resolution of
8 \AA$\,$ or a resolving power  equal to 625 at 5000 \AA.  During both
nights observing conditions were clear.

\noindent
Overall  we observed uninterruptedly  for 5.1  hours during  the first
night  and for  5.3  hours  during the  second.   Before starting  the
scientific exposures  we obtained  bias, flat field,  calibration lamp
spectra and a telluric standard  star with the same instrumental setup
described above.

\noindent
We adopted  a slit position angle  equal to 72.62 degrees  in order to
place the target star and its  close-by companion on the slit and kept
this configuration for the entire observing runs.

\noindent
Exposure  time was  set  to  12 sec\footnote{For  the  first night  we
  adopted initially 30  sec integration time switching then  to 12 sec
  after the  first 40 images.}  with a  read out time equal  to 45 sec
during the  first night and  to 12 sec  during the second. This  was a
consequence  of windowing  the  detector during  the  second night  of
observation.   A   summary  of  the  observations  can   be  found  in
Table~\ref{tab:observations}.

\section{Data analysis}
\label{sec:data_analysis}

The  data reduction was  performed with  our own  customised software.
After  applying bias  correction and  flat-fielding we  determined the
extraction  region for both  the target  and comparison  star spectra.
Along the  dispersion direction ($x$  axis) the extraction  region was
fixed between 970 pix up to  almost the end of the detector (2060 pix)
that is between $\sim$4900\AA and $\sim$8000\AA.
Along  the  spatial  direction  we  considered  an  extraction  window
including both spectra  (which were separated by around  44 pix on the
detector) and an additional 30 pix region on the top and the bottom to
properly measure  the sky.  We then calculated  the spectral centroids
and the  full width half maximum  (FWHM) separately in  each column by
fitting a  bi-gaussian analytic  function with exponential  wings. The
PSF profile  is slightly asymmetric and  for this reason  we adopted a
bi-gaussian function for  the central core of the  PSF.  Moreover, the
PSF is also  more extended than a simple  gaussian and the exponential
wings provided a more  accurate approximation.  Both profiles were fit
simultaneously,  while the  sky was  first measured  in  the outermost
regions and then fixed throughout the process.

\subsection{Cosmic rays correction}
\label{subsec:cosmics}

To  remove the  cosmic rays  contaminating  our data,  we shifted  and
rescaled the  spatial profile  of each column  into an average  PSF by
means  of  the analytical  solution  determined  previously.  We  then
spline interpolated  the PSF at  1/10th pixel resolution  shifting and
rescaling backward this  empirical PSF to the position  of each column
profile.  We  integrated numerically the PSF  over 1 pix  to match the
detector resolution.  We measured  the scatter around the average PSF,
and took a  generose threshold of 20 times this  value with respect to
the  shifted and  rescaled PSF  to isolate  possible outliers  in the
column profiles.  Once an outlier was found we replaced its value with
the model  PSF value.   On average we  measured 1-2 cosmic  events per
image.

\subsection{Extraction}
\label{subsec:extraction}

The  flux  of each  star  was  extracted  from the  measured  centroid
position in  each column  into an  aperture of 12  px radius  by using
aperture   photometry.    The  aperture   radius   was  chosen   after
experimenting  different apertures  taking the  one that  provided the
highest precision in the final photometry.

\subsection{Wavelength drift correction}
\label{subsec:wavelength_drift_correction}

\begin{figure}
\center
\epsscale{.80}
\plotone{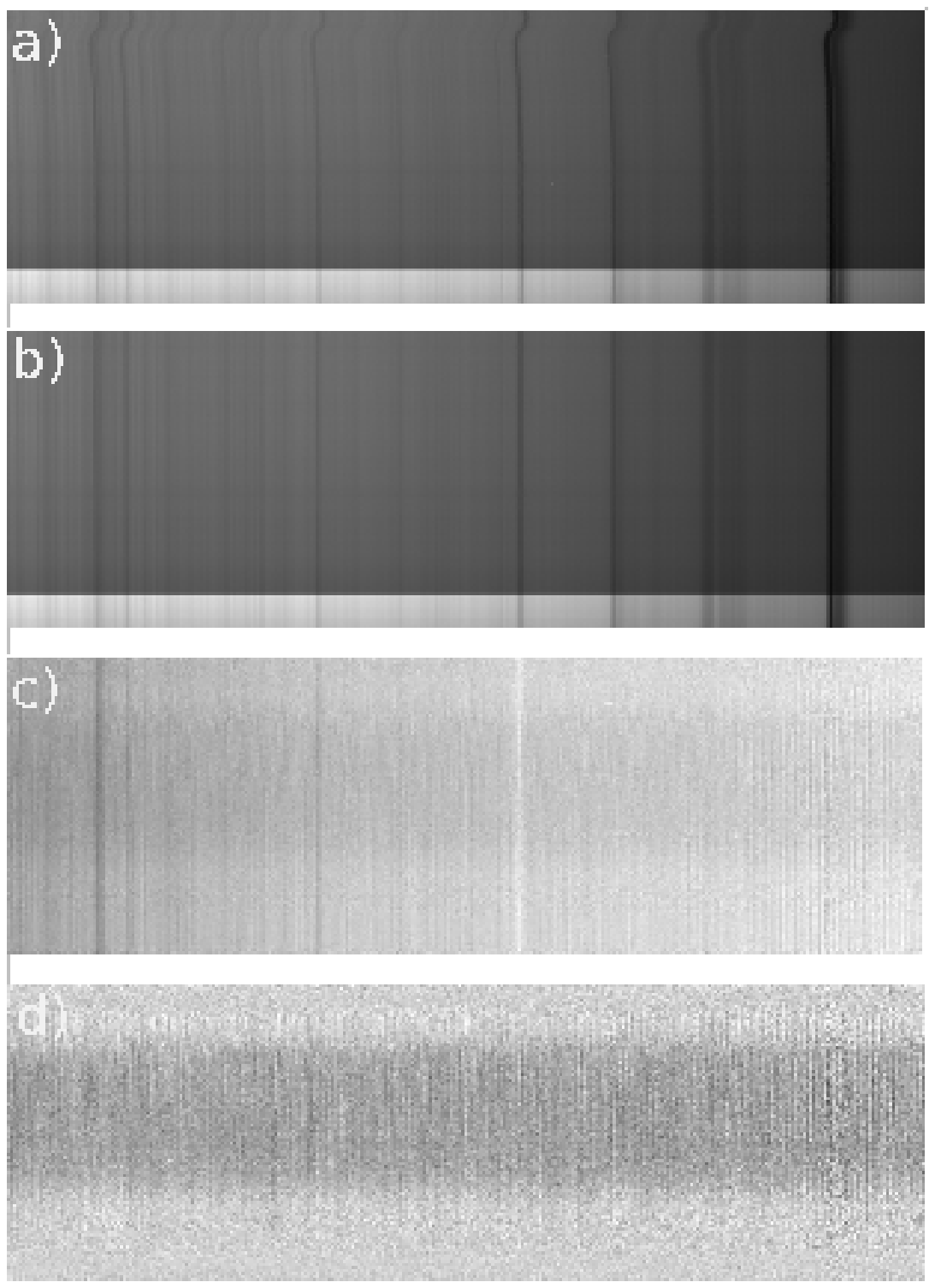}
    \caption{
             This Figure  highlights the major steps  in the reduction
             chain.  For each panel, dispersion direction is along the
             x  axis. Each line  represents a  spectrum obtained  at a
             given epoch. Spectra are chronologically ordered from the
             bottom to  the top of  each panel.  Data are  relative to
             the first  night of observation. a):  Observed spectra of
             the  target star b)  spectra are  re-aligned to  the same
             wavelenght  reference system  c) residual  spectrum after
             dividing  each  target   spectrum  to  the  correspondent
             comparison   star  spectrum  d)   differential atmospheric  
             extinction correction.   
             \label{fig:reduction}
            }
\end{figure}

After  extracting   the  flux   from  each  spectrum   we  constructed
bidimensional maps of dispersion versus time flux variations. In these
maps each line represents a spectrum of the target (or the comparison)
at a  given epoch.  In  Fig.~\ref{fig:reduction} (top panel),  we show
the target map obtained just after flux extraction for the first night
of  observation.  As  is  evident from  the  deepest absorption  lines
visible  in the  spectrum, the  spectra shifted  along  the dispersion
direction  during the  night.  
The maximum shift was equal to 1.7 arcsec (with respect to the initial
position)  during the first  night and  equal to  4 arcsec  during the
second. In both cases both the star and the comparison remained always
well within the slit thanks to the adoption of a large aperture.
Before lightcurve construction it is important to realign the spectra
and bring  them to  the same  reference system.  We  used the  out of
transit measurements  to construct an average spectrum  of the target
and of the comparison star, which was obtained by scaling the spectra
at mid-wavelength and  averaging their profiles. We used  then a deep
absorption  line to obtain  a first  guess of  the shift  between the
single spectra  and the average  spectrum from the difference  of the
minimum.  Then  we cross-correlated the  splined interpolated average
spectrum to  each single spectrum over  a region of 2  pix around the
initial guess, and at a step  of 1/10th of a pixel. Finally we spline
interpolated the orginal spectrum at half of a pixel and shifted them
accordingly  to  the  best  value   of  the  shift  obtained  by  the
cross-correlation.   The result  is shown  in Fig~\ref{fig:reduction}
(second   panel  from   the   top).   This   procedure  was   applied
independently both for the target and comparison spectra.

\subsection{Normalization to the comparison spectrum}
\label{subsec:normalization_to_the_comparison_spectrum}

We procedeed by dividing the target spectra by the comparison spectra,
obtaining the flux ratio of the two. This division was applied to each
pixel  of  the  bidimensional  maps  reported above.   The  result  is
illustrated  in Fig.~\ref{fig:reduction} (third  panel from  the top).
All but the some absorption lines in the original spectra disappear by
forming  the  flux  ratios,  demonstrating the  effectiveness  of  the
adoption of the comparison star.   The horizontal dark band visible in
all images corresponds to the transit event.

\subsection{Differential atmospheric extinction correction}
\label{subsec:atmospheric_extinction_correction}

One  of  the big  advantages  of  differential spectrophotometry  over
normal differential photometry is  the possibility to model and remove
more accurately the effect of  our own atmosphere.  The flux ratio map
presented  above displays both  a continuous  atmospheric differential
extintion  between  the  target  and  the comparison,  and  also  some
localized differential absorption.  To  further improve the result, we
then   applied  a   wavelength   dependent  differential   atmospheric
extinction  correction to  each single  detector column.   We  adopt a
linear model for  the flux variation as a function  of airmass and fit
it to each column of the flux  ratio maps by using only out of transit
data.   We then  normalized each  column by  the model  solution.  The
result is shown in Fig~\ref{fig:reduction} (fourth panel from the top)
and demonstrates that also  the residual contamination is well removed
by this approach.

\subsection{Uncertainties}
\label{subsec:uncertainties}

To each  flux measurement  we associated an  uncertainty based  on the
following theoretical formula:

\begin{equation}
\rm\frac{\Delta\,F\,}{F}=\,\sqrt{\sigma F_{0}^2+\sigma F_{s}^2}
\end{equation}

\noindent
with

\begin{equation}
\rm\sigma F_{0}=\frac{\sqrt{F+RON^2+\frac{\sigma_{sky}^2}{N_{sky}}}}{F}
\end{equation}

\begin{equation}
\rm\sigma F_{s}=0.09\,\frac{D^{-\frac{2}{3}}\,A^{1.75}\,\rm e^{-h/8}}{\sqrt{2\,T_{exp}}}
\end{equation}

\noindent
where F corresponds  to photon counts within the  aperture, RON is the
detector  read  out  noise,   $\rm\sigma  F_{0}$  is  the  photon  and
instrumental noise,  $\rm\sigma_{sky}$ is the uncertainty  in the mean
sky  determination (in  photons) and  $\rm N_{sky}$  is the  number of
pixels from  which the sky  was determined.  $\rm\sigma F_{s}$  is the
scintillation noise  which we estimated using the  formula reported in
Gilliland et al.~(1993)  based on the work of  Young et al.~(1967) and
where D is the telescope diameter in cm, A the airmass, h the altitude
of  the  observatory in  km  and  T$\rm_{exp}$  the exposure  time  in
seconds.

\noindent
Eq.~1  was  used to  construct  dispersion  versus time  bidimensional
uncertainty   maps   which  were   associated   with  the   scientific
measurements described in previous sections.

\subsection{Estimates}

\noindent
During our observations  we  acquired  around 9.7$\times$10$^7$  total
photons per exposure   for the comparison star and   5.6$\times$10$^7$
photons for the  target. Pure photon and  detector noise would imply a
precision of 0.013$\%$  per measurement, scintillation calculated from
the  above   formula would  add  the non-negligible contribution  of
0.027$\%$   at   low  airmasses,  depending    quite  sensibly  on the
airmass. We can expect  a total error  budget of around 0.03$\%$, that
is around  300  ppm.  The resulting   precision on the   transit depth
determination can be  estimated by considering  the error  on the mean
levels of the in and out  of transits measurements.  For the first and
second night we obtain an estimate of around 0.003$\%$ and 0.002$\%$.

\noindent
Assuming as in N14 that a scale height of HAT-P-1b is equal to 414 km,
this corresponds  to a required  photometric precision in  the transit
depth determination equal to 0.012$\%$ (or around 0.0005 in the radius
ratio  determination).  The  error  estimate reported  above does  not
account  for  some other  complications  present  in  the transit  fit
process, like for  example the necessity to model  the limb darkening.
What  it  demonstrates, however,  is  that  photon  noise is  not  the
limiting factor  in this experiment.  Even  rebinning the measurements
into few spectral  bins we should be able  to remain confortably below
the  limit imposed  by one  scale height.   The real  challange  is to
maintain  the accuracy  of  the  measurements down  to  this level  of
precision. For ground based observations this also means to accurately
remove the contribution of our own atmosphere.

\subsection{Contamination}
\label{subsec:contamination}
\begin{figure}
\center
\plotone{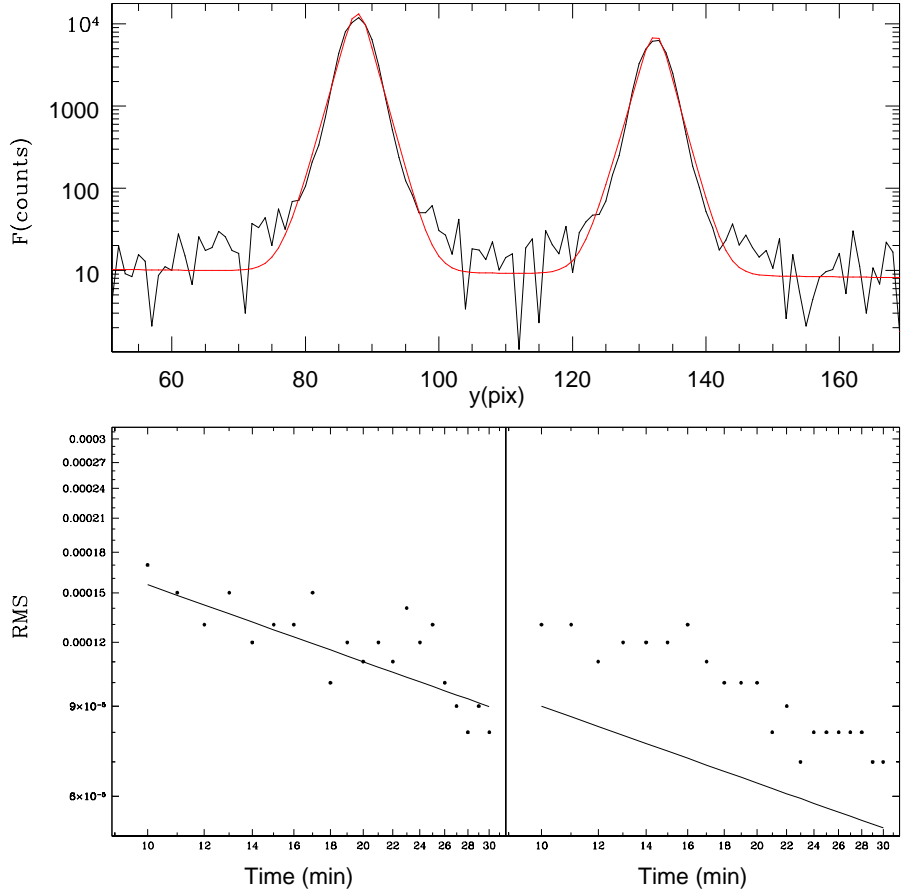}
    \caption{
             Top  panel:   spatial  cut  of  a   typical  spectrum  at
             mid-wavelentgth showing both  the comparison star and the
             target.   The red  continuous line  denotes  our best-fit
             model. Note the logarithmic  scale on the y-axis.  Bottom
             panels:   theoretical  RMS   (continuous   line)  against
             observed   RMS  (points)  as   a  function   of  temporal
             binning. On  the right white lightcurve  results from the
             first night, and on the left for the second night.
             \label{fig:contamination}
            }
\end{figure}

In Figure~\ref{fig:contamination} (top panel),  we show the profile of
the  comparison  and  target  star  along the  spatial  direction,  as
observed  at mid-wavelength for  a typical  spectrum.  The  y-scale is
logarithmic to  enhace the visibility  of the PSF wings.   This Figure
demonstrates that  the two objects  are well separated in  our images.
The  aperture radius  we  adopted  is around  4.6  times the  measured
fwhm. The centroid  of the target star is distant  around 17 fwhm from
the  centroid of the  comparison.  Considering  our best  fit analytic
model (denoted in red in the figure) this implies that the flux of the
comparison   integrated  across  the   target  aperture   is  entirely
negligible (nominally at a level of 10$^{-12}$ the target flux).

\subsection{Masking}
\label{subsec:masking}

For some  isolated measurements we noted that  errors were appreciably
different from the average.  These situations were probably associated
with  image artifacts and/or  cosmics not  perfectly corrected  by our
automatic  pipeline. To  mask these  regions we  adopted  an iterative
5-$sigma$ clipping algorithm measuring the average uncertainty and the
scatter  inside   a  10px  box  radius  window   scanning  the  entire
dispersion-time domain.  Overall,  the fraction of measurements masked
this way was equal to $\sim$0.05$\%$ of the total number of bins.

\subsection{Wavelength calibration}
\label{subsec:wavelength_calibration}

We  used a  Ne+Hg lamp  obtained just  before starting  the scientific
sequence to  map pixels into  wavelength, in angstroms.  The  lamp was
observed with the  same grism, but with a smaller slit  width of 2 arcsec
to allow a better identification of the spectral lines.

\section{Lightcurve analysis}
\label{sec:lightcurve_analysis}

Lightcurves  were obtained by  calculating a  weighted average  of the
flux  ratios in  different spectral  channels.  At  first, to  fix the
global system parameters, we simply averaged the measurements over all
the  spectral  range of  our  spectra.   The  resulting lightcurve  is
usually referred to as $white$ lightcurve.

\subsection{White lightcurves}
\label{subsec:white_lightcurves}

White lightcurves  were produced  and analyzed independently  for each
night of  observation.   
We assumed a perturbed
transit model.  As long  as systematics are  small as compared  to the
transit it  is always  possible to express  the global  model (transit
with systematics) as a sum  of an unperturbed transit model
(F$\rm_{MA}$) which is given in this case by the Mandel \& Agol (2002)
formula  and a  correcting term  which is  express below  as  a linear
combination of different systematic components:

\begin{equation}
\rm F(t)=F_{MA}(t)+c_1\,\times\,Air(t)+c_2\times\,xc(t)+c_3\times\,s(t)+c_4
\end{equation}

\noindent
where  c$\rm_1$,c$_2$ and c$_3$  are three  parameters to  account for
residual flux  losses related to airmass (Air),  spectral shifts along
the  dispersion direction  (xc) and  seeing (s).   The  constant c$_4$
accounts for a photometric zero point offset.

\noindent
We applied  the Bayesan Information  Criterium (BIC) to  determine the
best model to  fit the data considering different  combinations of the
above  parameters.   For the  first  night  of  observation the  model
reported in Eq.~4  was the favoured one, for the  second night the BIC
analysis indicated that the seeing  dependence could be dropped and we
thereby fixed c$_3$ to zero.

\noindent
For   the   planet-to-star   normalized   distance  we   adopted   the
parameterization described in Montalto et al.~(2012) and therefore fit
a  maximum  of nine  free  parameters:  the  time of  transit  minimum
($T_0$),  the planet  to star  radius ratio  ($r$), the  total transit
duration  ($T_{14}$), the mean  stellar density  ($\rho_{\star}$), the
linear limb  darkeneing coefficient  ($g1$) and the  four coefficients
c$_1$, c$_2$, c$_3$ and c$_4$.

\noindent
We  fitted  each  lightcurve  with the  Levemberg-Marquardt  algorithm
(Press 1992) using the partial derivatives of the flux loss calculated
by  P\'al  (2008) as  a  function  of the  radius  ratio  $r$ and  the
normalized distance  $z$ to calculate the  differential corrections to
apply to each parameter at each iterative step.

\noindent
We  assumed  a  quadratic  limb  darkening  law  where  the  quadratic
coefficient ($g2$) was obtained (and thereby fixed) using the software
developed  by Espinoza  \& Jord\'an  (2015) adopting  the  ATLAS model
atmospheres  results and  plugging  in the  response  function of  the
TNG+LRS intrument.   The quadratic coefficient was  calculated in this
way  both  for  the  white  lightcurves and  for  the  different  band
lightcurves.  HAT-P-1A spectroscopic parameters were obtained from the
SWEET-Cat      catalog       (Santos      et      al.~2013):      $\rm
T_{eff}=(6076\,\pm\,27)\,K$,     $\rm     lg_{LC}(g)=4.40\,\pm\,0.01$,
[Fe/H]=$+0.21\,\pm\,0.03$ and $\rm v_t=(1.17\,\pm\,0.05)\,km\,s^{-1}$.
We  also  calculated  in  the  same  way  the  value  for  the  linear
coefficient  $g1$  and used  this  value  as  initial guess  for  this
parameter.   The  initial  guesses  for the  other  system  parameters
($T_{14}$, $r$, $\rho_{star}$) were  obtained from the analysis of N14
as reported in their Table~5,  whereas for the slope parameters c$_1$,
c$_2$, and c$_3$  we assumed simply zero as initial  guess and for the
constant c$_4$, instead, one.

\noindent
For the orbital period we first  adopted the value reported by N14 and
derived the time  of transit minimum T$_0$ for  our transits using the
Levemberg-Marquardt  algorithm  described  below.  We  therefore  used
transit timings reported in the literature to further refine the value
of the  reference transit time T$_0$  as well as the  orbital period P
adopting a linear ephemerides

\begin{equation}
\rm T_C(E)\,=\,T_0\,+\,E\,\times\,P,
\end{equation}

\noindent
where  T$\rm_C$(E) is  the time  of transit  minimum at  epoch  E.  We
obtained       T$\rm_0$=(2453979.9321$\pm$0.0003)       days       and
P=(4.46529961$\pm$0.00000071)  days  with   a  reduced  chi  equal  to
$\chi_r=1.35$. In Table~2 we report  the list of transit times we used
in this calculation.
 
\noindent
We used  the Levemberg-Marquardt algorithm  (Press 1992) to  obtain an
initial  solution for  the free  parameters.  For  each  iteration the
the reduced  $\chi_{red}$ of the fit was calculated

\begin{equation}
\rm\chi_{red}\,=\,\sqrt{\sum_{i=1}^{i=N}\frac{(O_i-F_i)^2}{N-N_{free}}}
\end{equation}

\noindent
where $\rm  O_i$ is  the    observed flux  corresponding to the   i-th
measurement,    $\rm F_i$ is the   model  calculated flux as described
above, $\rm N$ is the total number of measurements, and $\rm N_{free}$
the number  of   free parameters.  The  Levemberg-Marquardt  algorithm
delivered the best solution by means of $\rm \chi_{red}$ minimization.

\subsection{Correlated noise}
\label{s:correlated_noise}

To  each photometric measurement we  associated an error equivalent to
the error of the weighted average  over integrated passband as derived
from our uncertainty maps. These errors were rescaled so that the best
models gave  as  result $\rm  \chi^2$=1.   The presence  of correlated
noise in  the data limits  the precision of  the observations (Pont et
al.~2006).  We then compared  the  RMS of  the fit residuals  averaged
over M  bins with the  theoretical prescription valid for uncorrelated
white noise

\begin{equation}
\rm\sigma_M=\frac{\sigma}{\sqrt{N}}\,\sqrt{\frac{M}{M-1}}
\end{equation}

\noindent
where N  is the total number of  residuals and $\sigma$ is  the RMS of
all residuals.  We considered bin sizes comprised in between 10 min to
30 min.  The average ratio ($\beta$) of the measured dispersion to the
theoretical  value  was  taken  as  a measurement  of  the  degree  of
correlation  in  the data,  and  the  uncertainty  were then  globally
expanded by  this factor. We found  that the $\beta$  parameter was in
between 1.05 and 1.62.

In  Figure~\ref{fig:contamination}   (bottom  panels),  we   show  the
thoretical RMS  (continuous line) and  the observed RMS (points)  as a
function of the  binning timescale for the first  and the second night
white  lightcurves  (left  and  right  respectively).  In  both  cases
residuals are  scaling down quadratically with  rebinning time though,
over  the analyzed  timescales, the  second night  appears to  be more
affected    by   red   noise    ($\rm\beta=1.4$)   than    the   first
($\rm\beta=1.1$).

\subsection{Monte carlo analysis}
\label{subsec:mcmc}

Subsequently  we performed  a  Markov Chain  Monte  Carlo analysis  as
described  in Montalto et  al. (2012)  to refine  the estimate  of the
uncertainties.  We considered 200000 iteration steps and 10 chains for
each lightcurve, merging then  the resulting chains after dropping the
first 20$\%$ burn-in phase iterations.

\begin{figure}
\center
\plotone{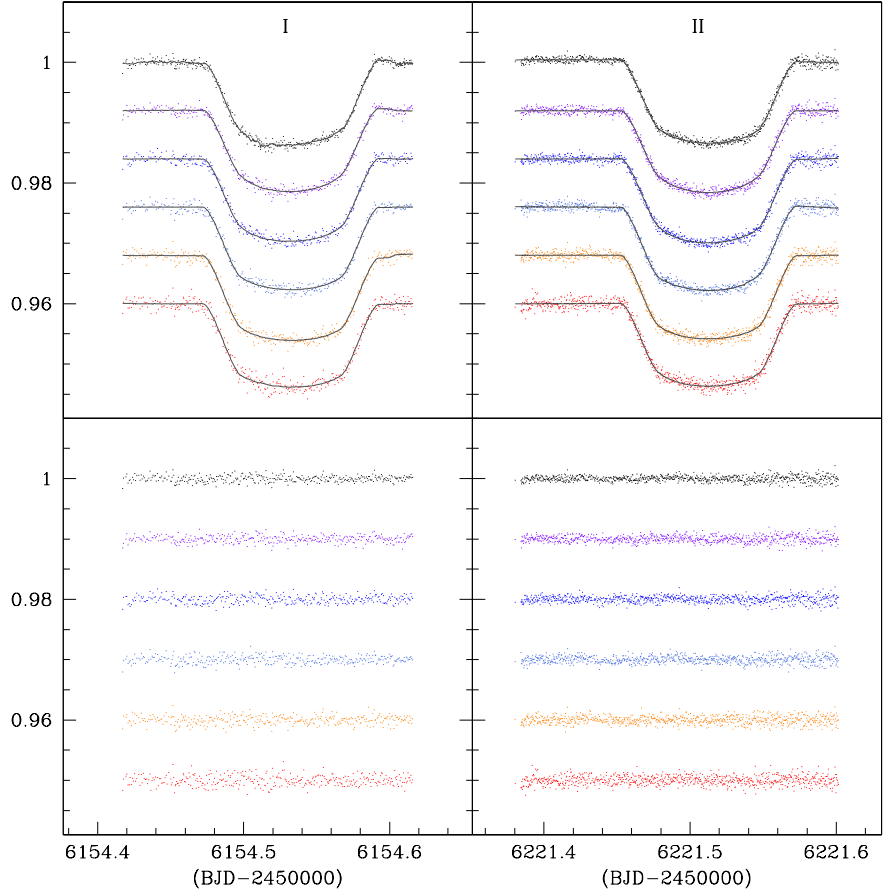}
    \caption{
             Top panels: lightcurves extracted from the data.
             The
             white lightcurve are on top followed by the color 
             lightcurves from the
             bluest to the reddest one (from top to bottom).  An offset
             is applied for  clarity. Bottom panels: Residuals  of the
             fit  ordered as in  the top panels. Left panels denote
             results  from  the first   night,  right panels  from the
             second.
             \label{fig:lightcurves}
            }
\end{figure}

\noindent
The  results  are illustrated  in  Fig.~\ref{fig:lightcurves} and  the
corresponding best-fit parameters are reported in Table~3.  The RMS of
the white lightcurves residuals is 525 ppm for the first night and 462
ppm for the second night.

\subsection{Multi-band lightcurves}
\label{subsec:multi_band_lightcurves}

We  experimented  various subdivisions  of  our  spectral window  into
different spectral bins trying to optimize the resolution of the final
spectrum   while  keeping   the   precision  of   the  transit   depth
determination  at a  level of  around  one scale  height.  Finally  we
adopted the a subdivision  into five $\sim$600\AA$\,$ wavelength bins.
A common mode systematic model  derived from the white lightcurves was
subtracted  from the  multi-band lightcurves.   The  analysis reported
above was then repeated for each lightcurve. In this case we fixed the
stellar density, time of transit minimum and total transit duration to
the  average  values  (of  the  two nights)  derived  from  the  white
lightcurve analysis.  This assumption is justified by the fact that we
do  not  expect these  parameters  to  be  wavelength dependent.   The
results  are   then  illustrated  in   Fig.~\ref{fig:lightcurves}  and
reported in Table~4.

\section{Results}
\label{sec:results}

\begin{figure}
\center
\epsscale{.80}
\plotone{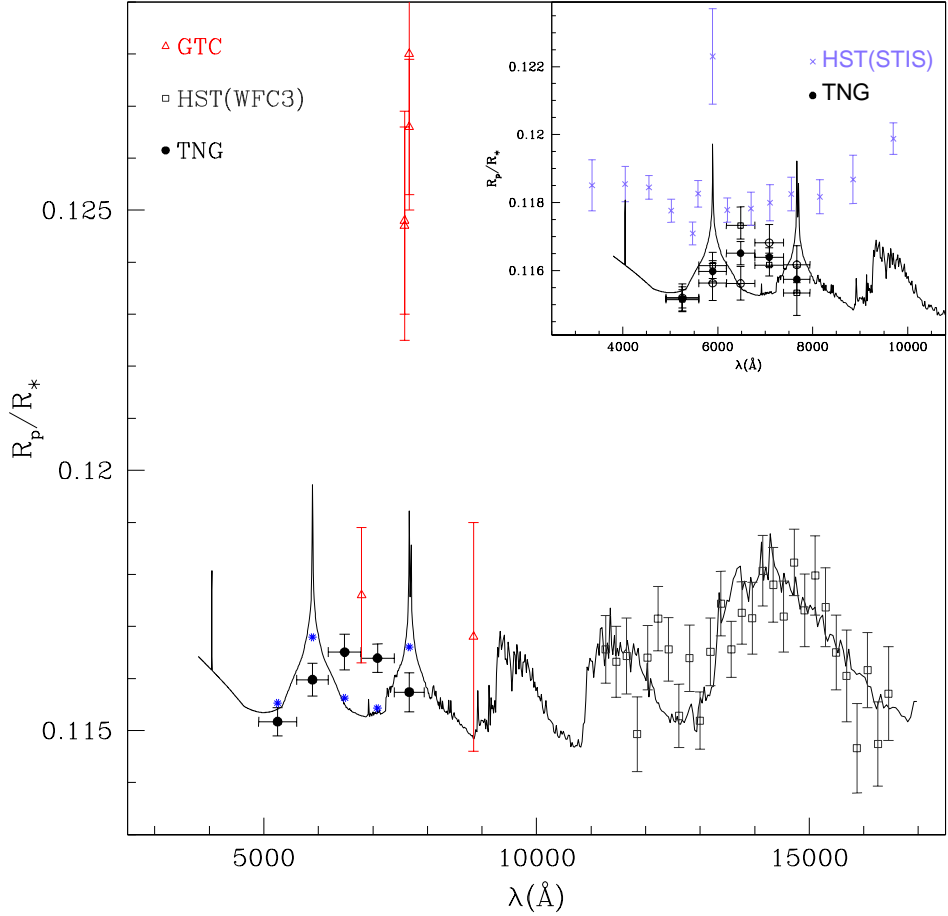}
    \caption{
             Comparison between different transmission spectroscopy
             measurements of HAT-P-1b and theoretical model predictions.
             In the main plot black circles denote TNG measurements
             presented in this work, grey boxes HST(WFC3) measurements
             (Wakeford et al. 2013), red triangles GTC measurements
             (Wilson et al. 2015). The model depicted is calculated
             for a solar abundance of Na and K. Blue asteriscs indicate
             model predictions averaged over TNG bandwidths. In the
             box on the top right, close-up view of the optical region
             displaying the HST(STIS) measurements (slate blue crosses,
             N14) and TNG measurements black circles. Open symbols denote
             single epoch TNG measurements (circles for the first epoch,
             boxes for the second). The same model displayed in the main
             figure is also presented.
             \label{fig:model_solar}
            }
\end{figure}

\begin{figure}
\center
\epsscale{.80}
\plotone{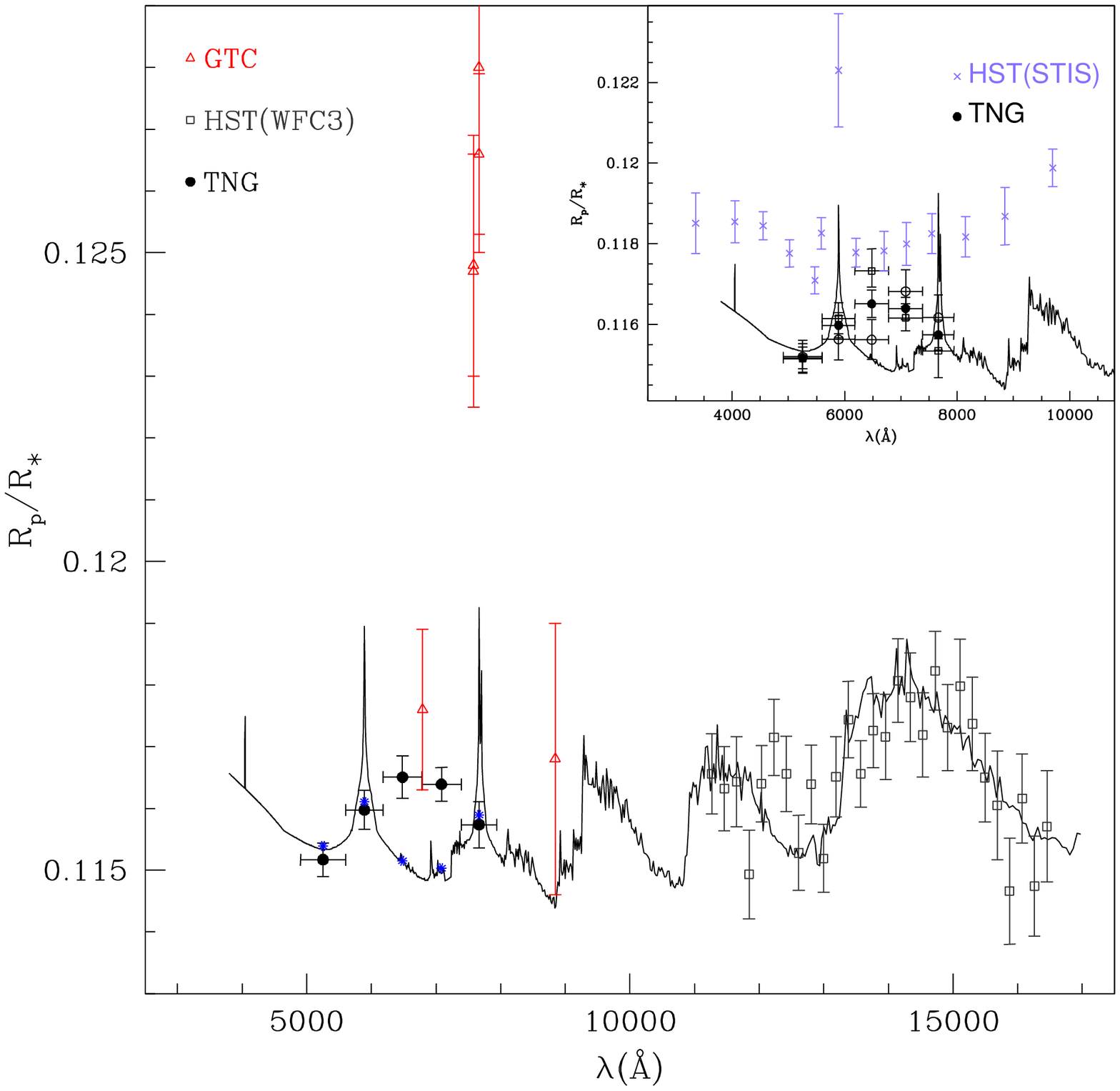}
    \caption{
             The  same  as Figure~3  with  a  model  calculated for  a
             subsolar (10$\rm\times$) abundance of Na and K.
             \label{fig:model_subsolar}
            }
\end{figure}

\begin{figure}
\center
\epsscale{.80}
\plotone{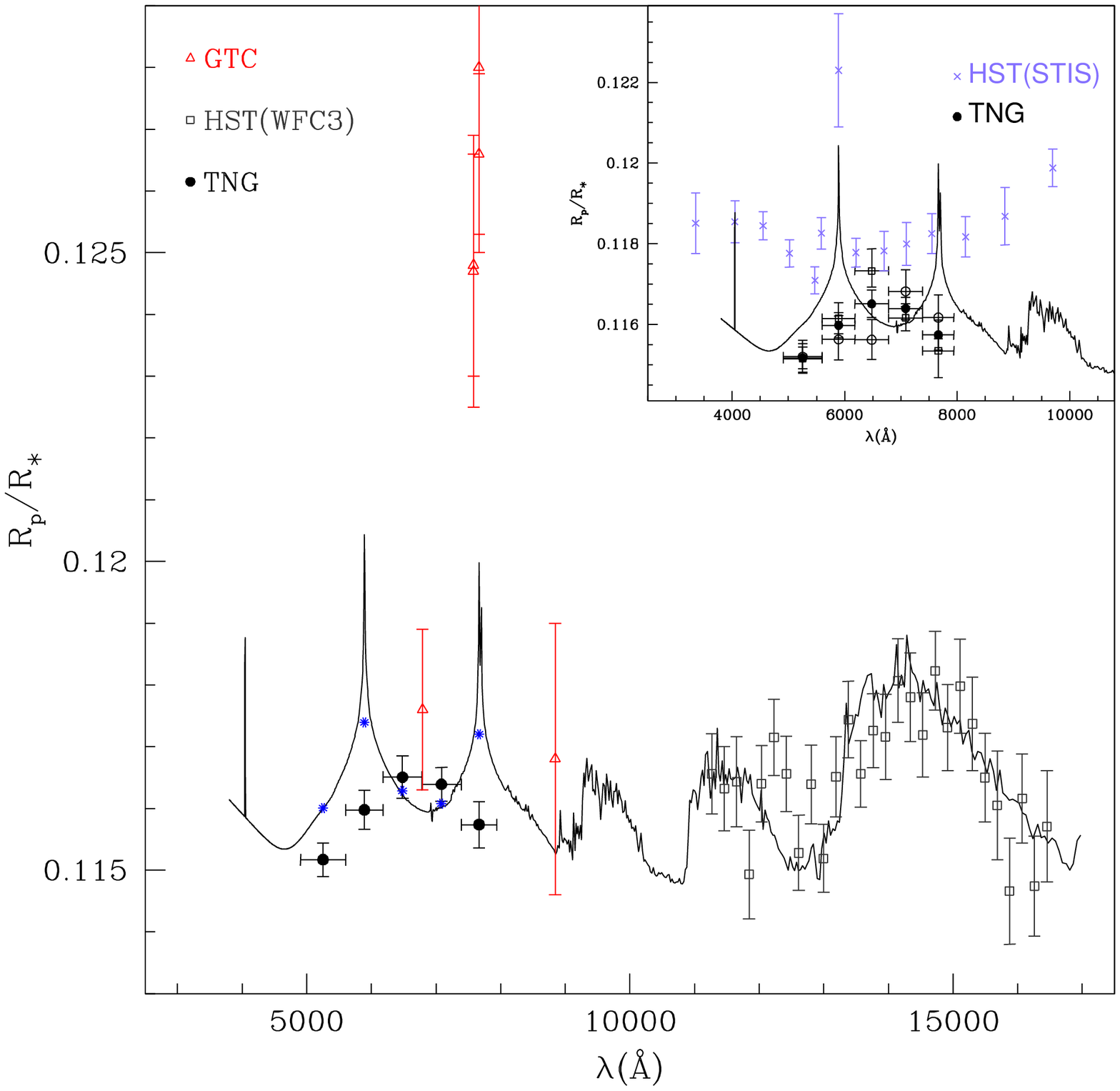}
    \caption{
             The  same  as Figure~3  with  a  model  calculated for  a
             supersolar (10$\rm\times$) abundance of Na and K.
             \label{fig:model_supersolar}
            }
\end{figure}

\noindent
In  Fig.~\ref{fig:model_solar}  to Fig.~\ref{fig:model_supersolar}  we
present the  results of our  analysis.  The black points  denoting our
final spectrum  were obtained  by taking the  weighted average  of the
results of both observing nights as reported in Table~5.

\noindent
A  fit with a  constant line  to our  data returns  a $\chi_{red}=$1.9
which may  appear a  bit high.  The  morphological structure  seems to
indicate  a single  peaked spectrum.   A  fit with  a quadratic  model
renstitutes a peak around 6650 \AA, close to the red edge of the third
bin.   The  measurement  with  the  lowest  radius  ratio  appears  to
correspond to  the bluest  bin.  Averaging the  measurements braketing
the peak in  the third and fourth bin and  forming the difference with
the bluest bin  we get a S/N equal to 3.7.   Further insights into the
characteristics of this spectrum  and its physical plausibility can be
obtained  by comparing these  results with  the predictions  of planet
atmosphere models.

\noindent
In  order  to  compute  the  synthetic spectra,  we  use  line-by-line
radiative transfer calculations  from 0.3 to 25 $\mu$m.   The model is
fully described in Iro et al.~(2005) and Iro \& Deming~(2010).

\noindent
As opacity sources we include the main molecular constituents: H$_2$O,
CO, CH$_4$, CO$_2$ , NH$_3$ (the  latter two are added with respect to
the previous references, taken  respectively from Rothman et al.~2009,
2010), Na,  K and TiO; Collision  Induced Absorption by  H$_2$ and He;
Rayleigh  diffusion and  H$^-$ bound-free  and H$_2^-$  free-free. The
current model does not account for clouds.

\noindent
In the  nominal model, we  assume the atmosphere is  in thermochemical
equilibrium with a solar abundance  of the elements. We generated also
two additional models with ten times higher and lower abundances of Na
and K with respect to the solar value.

\noindent
Planet and  star parameters are taken  from N14.  As  input heating we
use   a   stellar   spectrum   computed  from   Castelli   \&   Kurucz
(2003)\footnote{See   {\tt  ftp://ftp.stsci.edu/cdbs/grid/ck04models}}
for a G0V star with an effective temperature of 6000K.

\noindent
The  comparison of  the models  with  the global  set of  transmission
spectroscopy measurements  of HAT-P-1b acquired so  far is illustrated
in Fig.~3-Fig.~5.

\noindent
The  models  were  fitted  to  the  near infrared  data  of  Wakeford  et
al.~(2013)  and  then  simply  projected  in  the  optical.   We  then
calculated the reduced  $\chi_r$ of the models (averaged  over the TNG
bandwidths) with respect to the TNG measurements. The results indicate
that for all cases considered  a poor fit is obtained.  In particular,
the solar  abundance model  displayed in Fig.~3  implies $\chi_r=$2.9,
the super-solar  model (Fig.~5) $\chi_r=$3.4, and  the sub-solar model
(Fig.~4)  $\chi_r=$3.2. A closer  look at  these results  suggest that
clear models could be able to  fit a portion of the obtained spectrum,
but not the whole spectrum.  In particular, the subsolar model appears
to be able to perfectly reproduce  the blue edge cut-off of the sodium
line, as  well as  the sodium and  potassium region.   Restricting the
analysis to  these bins we  would obtain $\chi_r=$0.7 for  this model.
The discrepancy  in the region  in between Na  and K is  however quite
remarkable  (at the  level of  6.3  $\sigma$).  This  result could  be
interpreted as evidence  that the broad wings of  alkali metals in the
HAT-P-1b  atmosphere are narrower  than what  implied by  the standard
solar model.  Moreover,  it suggests that to be  able to reproduce the
entire spectrum it  is necessary to assume an  extra absorption in the
region in between Na and K.

\noindent
The plausibility  of this hypothesis can be  verified considering also
previous  observational results  and the  comparison  with theoretical
models which include the contribution of extra absorbers.

\noindent
In   particular  our   result   appears  similar   to   the  case   of
HD209458b. Des\'ert et al.~(2008) found  that for this planet the blue
side of the  sodium wing was well defined  but an increased absorption
was observed redward  of it.  They suggest that  the increased opacity
could be  explained by the presence  of a weak  absorption from TiO/VO
clouds, implying the presence of  a modest temperaure inversion in the
atmosphere.  In the  case of HAT-P-1b it seems  that the absorption is
more extended toward the red than in the case of HD209458b.
It is also  worth noting, however, that new  analysis of near infrared
data  appear  to  not confirm  early  claims  on  the existence  of  a
temperature inversion in HD209458b (Diamond-Lowe et al. 2014, Evans et
al. 2015).

\noindent
For  the  case  of  HAT-P-1b  secondary eclipse  measurements  in  the
IRAC/Spitzer  bands lead to  the conclusion  that HAT-P-1b  appears to
have a  modest thermal inversion  (Todorov et al. 2010).   Our optical
spectrum may therefore  support this conclusion, and it  could help to
identify the  optical counterpart of the opacity  source that produces
the observed infrared exess.  We  note that the same authors derived a
dayside  temperature of 1550$\pm$100  K (assuming  zero albedo  and no
heat redistribution).  This temperature appears quite low to attribute
our peak absorption  to TiO/VO clouds, given that  these compounds are
expected  to  condense  below 1600  K  but  we  note that  also  other
potential absorbers have been proposed (Burrows et al.~2007, Zahnle et
al.~2010).

\noindent
In conclusion,  our optical observations of HAT-P-1b  suggest that the
planet has a partially clear  atmosphere with a modest absorption from
clouds  (or other  absorbers) concentrated  in the  region  in between
6180-7400 \AA.

\subsection{Impact of model assumptions}
\label{model_assumptions}

In this  subsection we  investigate the dependence  of our  results on
several assumptions that have been considered during our analysis.

\noindent
First  of  all, as  reported  in  the  previous sections,  during  our
analysis   we   considered   a   semi-empirical  approach   to   model
limb-darkening and  adopted a quadratic  limb-darkening law. Quadratic
approximation  is usually sufficient  to describe  transit lightcurves
considering the actual precisions of observational data. However other
authors, like for  example N14, prefer to dopt  a four coefficient law
setting the coefficients to theoretical values. A four coefficient law
is expected to reproduce  more accurately the radial flux distribution
across the stellar  disk, in particular at the limb  (e.  g. Claret \&
Bloemen 2011).   We tried then to investigate  the possible dependence
of  our  results on  the  adopted  limb-darkening  law.  We  therefore
considered a  four coefficients limb-darkening law  and replicated the
procedure  reported  above  setting  the coefficients  to  the  values
obtained  by  using  the  software  of  Espinoza  \&  Jord\'an  (2015)
considering our white lightcurve passband and the response function of
our   instrumentation.    This   procedure   delivered    the   values
(0.5884,-0.0445,0.3287,-0.1984)   for   the   four   coefficients   in
increasing order  of power.  Additionally, in order  to reproduce more
closely the analysis of N14 we  also fixed the stellar density and the
transit duration to the exact values reported by N14. We note however,
that  the  stellar  density  we  derived  is  consistent  within  0.75
$\rm\sigma$  and  the transit  duration  within  1.8 $\rm\sigma$  with
respect to  the values reported by  N14.  The result  of this analysis
implies  a radius  ratio equal  to (0.1161$\pm$0.0003)  for  the first
night white lightcurve and equal to (0.1160$\pm$0.0004) for the second
night  white lightcurve.   This  test therefore  further supports  our
conclusion on the presence of a largely clear atmosphere in HAT-P-1b.

\noindent
We also  tested the adoption  of a different  mathematical formulation
for the systematic model analysis.  In particular, we considered as in
Stevenson  et  al.~(2014)  a  model  where the  systematic  terms  are
factorized rather  than summed to the unperturbed  transit model. Each
systematic term  is then assumed of the  form (1+$a\,\times\,$x) where
$a$ is a  modulating factor and x is  the normalized (mean subtracted)
variable   describing    a   given   systematic    (airmass,   seeing,
etc.). Expressing the model in  this way should reproduce more closely
the  physical  mechanisms underlying  the  systematic flux  variation.
Given the small amplitude of  systematics in our data this formulation
produced results which are fully consistent with the perturbed transit
model  analysis  described  above.    In  particular,  for  the  white
lightcurves  transit   depths  we  obtained   (0.1149$\pm$0.0008)  and
(0.1164$\pm$0.0007) for the first and second nights respectively, that
is well within our quoted uncertainties. This result then supports our
analysis and adopted formulation for systematics modeling.

\subsection{Discussion}
\label{sec:discussion}

\noindent
While we concluded  that clouds (or extra absorbers)  are necessary to
explain  the spectrum  of HAT-P-1b,  they overall  impact seems  to be
substantially smaller than  the one reported by N14,  as also shown in
the             top            right             corners            of
Fig.~3-Fig.~5.

\noindent
In the next subections,  we investigate three possible mechanisms that
could be at the origin  of the observed difference between our optical
results and the one presented in N14.

\subsection{Stellar activity}
\label{stellar_activity}

Stellar activity is a known  phenomenon that can produce radius ratios
variations (e.  g. Oshagh et  al.~2014).  HAT-P-1b is not  regarded an
active  star.  A very  low chromospheric  activity in  the CaII  and K
lines has been measured (Bakos  et al.~2007; Knutson et al.~2010) with
log(R'$\rm_{HK}$)=-4.984.  Transit  follow-up analysis (including this
work) did not  show evidence of spot crossing  events.  In addition to
that, photometric monitoring was  conducted at the Liverpool Telescope
between 2012 May  4 and 2012 December 13, covering  all the HST visits
during which HAT-P-1b  was observed as presented in  N14.  Because our
measurements  have  been acquired  in  August  and  October 2012,  the
photometric constraints derived by the authors apply also to our case.
The  Liverpool  Telescope follow  up  implied  that  HAT-P-1A was  not
variable during  the observing  period down to  a precision  of around
0.2$\%$ as we can infer from N14 (Figure 2).  Visit 20 of HST observed
with the G750  grism (overlapping our spectral window)  and produced a
radius  ratio  equal  to  R$\rm_p$/R$\rm_s$=0.11808$\pm$0.00034.   Our
average   radius    ratio   is   R$\rm_p$/R$\rm_s$=0.11590$\pm$0.0005.
Squaring these two radius  ratio estimates and taking their difference
we  obtain  an  expected   level  of  photometric  variability  around
0.05$\%$.   Such  a  value  is  around four  times  smaller  than  the
precision reached by the  Liverpool Telescope.  This calculation shows
that  probably  it  is   premature  to  completely  rule  out  stellar
variability  as a  potential  source of  systematic  variation of  the
radius ratio.  In particular it results that the observed radius ratio
difference could be produced if a sub-millimag variability is present.

\subsection{Exo-weather}
\label{exo-weather}

Although  in   the  previous  section  we  argued   that  the  current
photometric constraints are not very  tight to completely rule out the
stellar variability  hypothesis, the level of  variability required to
explain the radius variation appears indeed quite small.  Assuming for
HAT-P-1b an atmospheric scale height of 414 km as in N14, this implies
that the observed difference would amount to around 4.4 scale heights.
Atmospheric effects produced in the atmosphere of the exoplanet may be
present over a range of a  few scale heights so that it is interesting
to  speculate  on  the  possibility  that the  observed  radius  ratio
difference may be produced by a  mechanism which has its origin in the
atmosphere of the planet rather than in the one of its host star.  The
major factor  expected to affect  exo-planetary spectra is  related to
the presence or absence of clouds in their atmospheres.  The impact of
clouds  depends  critically  on  the  altitude where  the  clouds  sit
(Ehrenreich  et  al. 2006).   High  altitude  clouds  are expected  to
produce  the most  relevant effect  given  that they  prevent us  from
observing the lower level of the atmosphere, where high pressure wings
of alkali metals are expected to form.  This means that low resolution
spectroscopic   measurements   will   essentially   produce   a   flat
spectrum. If  the clouds sit at  a lower altitude their  impact on the
overall  spectrum is smaller,  especially if  they are  not completely
thick, and  atmospheric features may become comparably  more easier to
detect at low  resolution, given that higher pressure  levels are more
accessible.  In order to  reconcile these considerations with HAT-P-1b
observations we  should therefore  hypothesize that in  its atmosphere
there are  at least  two cloud  layers. The cloud  layer of  N14 would
represent the top layer, whereas  the one that seems to best reproduce
our  observations would  be the  bottom layer.   If the  top  layer is
present  we should  expect to  observe a  larger radius  ratio  and an
overall  flat spectrum, if  it is  absent the  radius ratio  should be
smaller and the spectrum, if  the bottom cloud layer is not completely
thick, may display  features like the blue edge  cut-off of the sodium
high pressure wing predicted by models.  This double cloud layer model
could therefore  at least qualitatively  explain the variation  of the
radius  ratio and  of  the morphological  structure  of the  spectrum.
However,  another  important  assumption   needs  to  be  made  to  be
consistent with  the observations.  Considering  that all observations
relevant  for this work  have been  acquired within  a few  months, we
should admit that exo-weather (in particular the conditions on the top
cloud layer) may evolve on timescales of weeks to months.

\subsection{Instrumental systematic}
\label{s:instrumental_systematic}

Less fascinating than the  previous hypothesis is the possibility that
the  observed  radius  ratio  difference  could  be  also  due  to  an
instrumental systematic.   Analyzing the results  presented in Fig.3-5
it  is probably  wise to  recall  that we  are comparing  observations
obtained  with  different instruments  under  a  variety of  observing
conditions.  The most relevant difference between our analysis and the
one  conducted  by  N14  is  likely  the  fact  that  to  correct  for
systematics we  used the  comparison star HAT-P-1B.   For ground-based
observations  differential  analysis is  simply  mandatory, while  for
space  based  observations  pure  instrumental photometry  along  with
refined detrending techniques achieves better performances than on the
ground,  but important  systematics are  present and  the  adoption of
comparison stars largely reduce their impact.  In a recent work B\'eky
et  al.~(2013)   studied  the  albedo  of  HAT-P-1b   using  the  STIS
spectrograph on  board of HST.   Performing a comparative  analysis of
the results considering the adoption or not of the comparison star the
authors  concluded  that   using  the  comparison  largely  simplified
detrending procedures eliminating important biases on the results.

\section{Conclusions}
\label{sec:conclusions}

In this  work we tested the  differential spectrophotometry technique
to  probe the  atmosphere of  the exoplanet  HAT-P-1b.  We  used  as a
calibrator  the  close-by   stellar  companion  HAT-P-1B,  of  similar
magnitude and  color than  the target star  HAT-P-1A. The  two objects
have been  simultaneously observed  by placing them  on the  slit.  We
used  the DOLORES  spectrograph at  the  TNG observing  two nights  on
August  14, 2012  and October  20,  2012.  We  achieved a  photometric
precision of 525 ppm with a  sampling of 52 sec during the first night
and  462 ppm with  a sampling  of 24  sec during  the second.  We also
demonstrate  the  possibility to  model  the  transit (including  limb
darkening)  with  a  relatively   simple  model.  The  result  of  our
observations  imply an  average  of R$\rm_p$/R$_*$=(0.1159$\pm$0.0005)
lower than previously estimated.  We observed a single peaked spectrum
(at a  3.7$\sigma$ level) and deduced  the presence of  a blue cut-off
corresponding to the blue edge of the sodium broad absorption wing and
an increased  opacity in  the region in  between sodium  and potassium
(6180-7400  $\rm\AA$). We  explain this  result as  the evidence  of a
partially   clear  atmosphere.    We  confirm   the  presence   of  an
extra-absorber in  the atmosphere of  HAT-P-1b as inferred by  N14 but
its impact on the overall spectrum seems to be more modest in our data
than what implied by previous observations.

\acknowledgments

This  work  was supported  by  Funda\c{c}\~ao  para  a Ci\^encia  e  a
Tecnologia  (FCT) through the  research grant  UID/FIS/04434/2013.  MM
acknowledges   the   support   from   FCT  through   the   grant   and
SFRH/BDP/71230/2010 and  kindly aknowledges  Luca Di Fabrizio  and the
TNG  staff for their  support during  the observations  and insightful
comments on the technical  specifications of the instrumentation.  The
anonymous  referee is  also aknowledged  for the  useful  comments and
suggestions.  PF and NCS  acknowledge support by Funda\c{c}\~ao para a
Ci\^encia e  a Tecnologia (FCT) through Investigador  FCT contracts of
reference IF/01037/2013 and  IF/00169/2012, respectively, and POPH/FSE
(EC) by  FEDER funding through  the program ``Programa  Operacional de
Factores  de  Competitividade  -  COMPETE''. PF  further  acknowledges
support from Funda\c{c}\~ao  para a Ci\^encia e a  Tecnologia (FCT) in
the    form     of    an    exploratory     project    of    reference
IF/01037/2013CP1191/CT0001. R.A. acknowledges  the Spanish Ministry of
Economy and  Competitiveness (MINECO) for the  financial support under
the Ram\'on y Cajal program RYC-2010-06519.

\clearpage

\clearpage

\clearpage

\clearpage

\clearpage

\clearpage

\begin{table}
\centering
\caption{
Observations
\label{tab:observations}
}
\begin{center}
\begin{tabular}{c c c c c}
\hline
\hline
Date & Epoch & Texp & Airmass range & N. spectra \\
\hline
14/08/2012 & 487 & 12 & 1.9-1.0 & 350 \\
20/10/2012 & 502 & 12 & 1.0-1.6 & 809 \\
\hline
\end{tabular}
\end{center}
\end{table}

\clearpage

\begin{deluxetable}{cccc}
\tabletypesize{\scriptsize}
\rotate
\centering
\tablecaption{Epochs used in the ephemerides calculation}
\tablewidth{0pt}
\tablehead{
\colhead{Epoch}  & \colhead{Time of transit minimum (BJD) } & \colhead{Error (BJD)} & Reference\tablenotemark{a}
}
\startdata
    0  &     3979.93071    &       0.00069     &     2 \\
    1  &     3984.39780    &       0.00900     &     1 \\
    2  &     3988.86274    &       0.00076     &     2 \\
    4  &     3997.79277    &       0.00054     &     2 \\
    4  &     3997.79425    &       0.00047     &     2 \\
    6  &     4006.72403    &       0.00059     &     2 \\
    8  &     4015.65410    &       0.00110     &     2 \\ 
   20  &     4069.23870    &       0.00290     &     2 \\
   86  &     4363.94677    &       0.00091     &     3 \\
   90  &     4381.80920    &       0.00130     &     3 \\
  469  &     6074.15737    &       0.00018     &     4 \\
  470  &     6078.62307    &       0.00018     &     4 \\
  478  &     6114.34537    &       0.00020     &     4 \\
  487  &     6154.53253    &       0.00015     &     5 \\
  495  &     6190.25542    &       0.00018     &     4 \\
  502  &     6221.51261    &       0.00019     &     5 \\
\enddata
\tablenotetext{a}{Reference: 1 - Bakos et al. (2007); 2 - Winn et al. (2007)
3 - Johnson et al. (2008); 4 - Nikolov et al. (2014); 
5 - This Work}
\end{deluxetable}

\begin{deluxetable}{cccccccccc}
\tabletypesize{\tiny}
\rotate
\tablecaption{White lightcurves best fit parameters\tablenotemark{a}}
\tablewidth{0pt}
\tablehead{
\colhead{$\rm R_p/R_{\star}$} & \colhead{T$\rm_0$(days)} & \colhead{T$\rm_{14}$(days)} & \colhead{$\rm\rho$(g$\,$cm$^{-3}$)} & \colhead{g1} & \colhead{g2} & \colhead{c$\rm_1$} & \colhead{c$\rm_2$} & \colhead{c$\rm_3$} & \colhead{c$\rm_4$} 
}
\startdata
$ 0.1153_{-0.0011}^{+0.0006}$    & $6154.53253_{-0.00013}^{+0.00018}$ &  $0.11984_{-0.00094}^{+0.00078}$  & $0.981_{-0.051}^{+0.057}$ & $ 0.420_{-0.045}^{+0.048}$ & $0.2875$ &  $-0.00024_{-0.00020}^{+0.00010}$ & $-0.000047_{-0.000018}^{+0.000015}$ & $-0.00114_{-0.00019}^{+0.00017}$ & $ 1.00195_{-0.00037}^{+0.00031}$ \\
$ 0.11633_{-0.00065}^{+0.00079}$ & $6221.51261_{-0.00017}^{+0.00022}$ &  $0.12007_{-0.00056}^{+0.00062}$ & $0.916_{-0.043}^{+0.028}$ & $ 0.393_{-0.045}^{+0.038}$ & 0.2875 & $-0.0008_{-0.0015}^{+0.0002}$ & $ 0.000000_{-0.000016}^{+0.000036}$ & $0$ & $ 1.0013_{-0.0005}^{+0.0013}$ \\
\enddata
\tablenotetext{a}{Values without uncertainty have been fixed during the fit}
\end{deluxetable}

\clearpage

\begin{deluxetable}{cccccccc}
\tabletypesize{\scriptsize}
\rotate
\centering
\tablecaption{Spectral band lightcurves best fit parameters}
\tablewidth{0pt}
\tablehead{
\colhead{Band ($\rm\AA$)} & \colhead{$\rm R_p/R_{\star}$} & \colhead{g1} & \colhead{g2} & \colhead{c$\rm_1$} & \colhead{c$\rm_2$} & \colhead{c$\rm_3$} & \colhead{c$\rm_4$} 
}
\startdata
$4906$-$5600$ & $ 0.11520_{-0.00041}^{+0.00038}$ & $ 0.495_{-0.022}^{+0.014}$ & 0.2644 & $-0.00015_{-0.00022}^{+0.00010}$ & $-0.000032_{-0.000015}^{+0.000013}$ & $-0.00013_{-0.00022}^{+0.00015}$ & $ 1.00048_{-0.00038}^{+0.00036}$ \\
$5600$-$6183$ & $ 0.11562_{-0.00059}^{+0.00051}$ & $ 0.414_{-0.035}^{+0.019}$ & 0.2917 & $-0.00005_{-0.00027}^{+0.00016}$ & $-0.000005_{-0.000022}^{+0.000018}$ & $-0.00019_{-0.00028}^{+0.00024}$ & $ 1.00034_{-0.00057}^{+0.00045}$ \\
$6183$-$6780$ & $ 0.11562_{-0.00050}^{+0.00049}$ & $ 0.335_{-0.032}^{+0.021}$ & 0.3045 & $ 0.00005_{-0.00024}^{+0.00016}$ & $ 0.000005_{-0.000023}^{+0.000014}$ & $ 0.00005_{-0.00025}^{+0.00025}$ & $ 0.99985_{-0.00060}^{+0.00037}$ \\
$6780$-$7390$ & $ 0.11682_{-0.00053}^{+0.00038}$ & $ 0.315_{-0.030}^{+0.019}$ & 0.2945 & $ 0.00009_{-0.00019}^{+0.00018}$ & $ 0.000057_{-0.000020}^{+0.000015}$ & $-0.00010_{-0.00027}^{+0.00021}$ & $ 0.99993_{-0.00057}^{+0.00037}$ \\
$7390$-$7942$ & $ 0.11617_{-0.00056}^{+0.00052}$ & $ 0.302_{-0.034}^{+0.025}$ & 0.2933 & $ 0.00011_{-0.00026}^{+0.00017}$ & $ 0.000009_{-0.000022}^{+0.000019}$ & $ 0.00022_{-0.00039}^{+0.00021}$ & $ 0.99950_{-0.00057}^{+0.00057}$ \\
\tableline
$4906$-$5600$ & $ 0.11514_{-0.00038}^{+0.00036}$ & $ 0.519_{-0.017}^{+0.011}$ & 0.2644 & $ 0.00003_{-0.00064}^{+0.00036}$ & $ 0.000001_{-0.000014}^{+0.000014}$ & $ 0$ & $ 0.99995_{-0.00052}^{+0.00052}$ \\
$5600$-$6183$ & $ 0.11615_{-0.00038}^{+0.00038}$ & $ 0.443_{-0.018}^{+0.011}$ & 0.2917 & $ 0.00105_{-0.00075}^{+0.00037}$ & $-0.000033_{-0.000014}^{+0.000016}$ & $ 0$ & $ 0.99891_{-0.00057}^{+0.00061}$ \\
$6183$-$6780$ & $ 0.11733_{-0.00055}^{+0.00040}$ & $ 0.390_{-0.020}^{+0.017}$ & 0.3045 & $-0.00182_{-0.00077}^{+0.00058}$ & $ 0.000055_{-0.000020}^{+0.000016}$ & $ 0$ & $ 1.00190_{-0.00082}^{+0.00059}$ \\
$6780$-$7390$ & $ 0.11616_{-0.00035}^{+0.00032}$ & $ 0.345_{-0.015}^{+0.014}$ & 0.2945 & $ 0.00054_{-0.00048}^{+0.00050}$ & $-0.000019_{-0.000017}^{+0.000010}$ & $ 0$ & $ 0.99946_{-0.00067}^{+0.00035}$ \\
$7390$-$7942$ & $ 0.11534_{-0.00037}^{+0.00067}$ & $ 0.370_{-0.027}^{+0.014}$ & 0.2933 & $ 0.0007_{-0.0013}^{+0.0002}$    & $-0.000023_{-0.000012}^{+0.000029}$ & $ 0$ & $ 0.9993_{-0.0005}^{+0.0011}$ \\
\enddata
\end{deluxetable}

\clearpage

\begin{deluxetable}{ccc}
\tabletypesize{\scriptsize}
\rotate
\centering
\tablecaption{Final radius ratio measurements}
\tablewidth{0pt}
\tablehead{
\colhead{Band ($\rm\AA$)}  & \colhead{$\rm R_p/R_{\star}$} & \colhead{$\rm \sigma(R_p/R_{\star})$} 
}
\startdata
$4906$-$5600$ & 0.1152 & 0.0003 \\
$5600$-$6183$ & 0.1160 & 0.0003 \\
$6183$-$6780$ & 0.1165 & 0.0003 \\
$6780$-$7390$ & 0.1164 & 0.0003 \\
$7390$-$7942$ & 0.1157 & 0.0004 \\
\enddata
\end{deluxetable}

\end{document}